# Apparent Correction to the Speed of Light in a Gravitational Potential

**J.D. Franson**
Physics Department, University of Maryland, Baltimore County, Baltimore, MD 21250

E-mail: jfranson@umbc.edu

**Abstract:** The effects of physical interactions are usually incorporated into the quantum theory by including the corresponding terms in the Hamiltonian. Here we consider the effects of including the gravitational potential energy of massive particles in the Hamiltonian of quantum electrodynamics. This results in a predicted correction to the speed of light that is proportional to the fine structure constant. The correction to the speed of light obtained in this way depends on the gravitational potential and not the gravitational field, which is not gauge invariant and presumably nonphysical. Nevertheless, the predicted results are in reasonable agreement with experimental observations from Supernova 1987a.

## 1. Introduction

One might suppose that the effects of a gravitational field on a quantum system could be described, at least to a first approximation, by including the gravitational potential $\Phi_G$ in the Hamiltonian. That approach has been successfully used, for example, to analyze the results of neutron [1] and atom [2-10] interferometer experiments in a gravitational field. Here we consider a model in which $\Phi_G$ is included in the Hamiltonian of quantum electrodynamics. As a result, virtual electron-positron pairs [11-16] have a gravitational potential energy that is the same as that of real particles. A straightforward calculation based on that assumption shows that the velocity of light in a gravitational potential would be reduced by an amount that is proportional to the fine structure constant $\alpha$.

The predicted correction to the speed of light depends on the gravitational potential and not the gravitational field, and it could be observed locally by comparing the velocity of photons and neutrinos, for example. As a result, the predicted correction to the speed of light is not gauge invariant. These results are also not equivalent to what would be obtained [17-19] from the currently-accepted generalization of the Dirac equation and quantum electrodynamics to curved spacetime [20-25]. The lack of gauge invariance and the disagreement with the generally-covariant Dirac equation both suggest that including the gravitational potential in the Hamiltonian must be nonphysical.

Nevertheless, the predicted correction to the speed of light from this simple model is in reasonable agreement with experimental observations from Supernova 1987a, where the first neutrinos arrived approximately 7.7 hours before the first photons [26]. There is no conventional explanation for how that could have occurred and the currently-accepted interpretation of the data is that the first burst of neutrinos must have been unrelated to the supernova [26], despite the fact that the probability of such an event having occurred at random is less than $10^{-4}$ [27]. The predicted correction to the speed of light, if correct, could explain this long-standing anomaly.

Quantum mechanics and general relativity are two of the most fundamental laws of physics. Quantum mechanics has been verified to very high precision by quantum electrodynamics experiments such as the measurement of the electron g-factor [28,29]. Experimental tests of general relativity are much more limited and many of the observed phenomena are consistent with other formalisms. As a result, there is currently a great deal of interest in performing high-precision tests [30] of general relativity using the properties of quantum systems, such as atom interferometers [2-10] superconductors [31-34], and photons [35,36]. The correction to the speed of light predicted here is closely related to the equivalence principle, as will be described below, and these results may provide additional motivation for experimental tests of general relativity, especially the equivalence principle.

Einstein was the first to predict that the velocity of light would be reduced by a gravitational potential [37]. According to general relativity [38-39], the speed of light $c$ as measured in a global reference frame is given by

$$c = c_0 \left( 1 + 2\frac{\Phi_G(\mathbf{r})}{c_0^2} \right), \tag{1}$$

where $c_0$ is the speed of light as measured in a local freely-falling reference frame. This reduction in the speed of light can be observed if a beam of light passes near a massive object such as the sun, as illustrated in Fig. 1. The transit time from a distant planet or satellite to Earth can be measured as a function of the distance $D$ of closest approach to the sun and then compared to the transit time expected at a velocity of $c_0$. The results from such experiments [40] are in excellent agreement with the prediction of Eq. (1). The deflection of starlight by a massive object can also be intuitively understood in this way. It should be noted that the velocity of light measured by a local observer will be independent of $\Phi_G$ and that the observable effects in this example are due to the spatial variations in $\Phi_G$.

The model considered here gives a correction to Eq. (1) that is proportional to the fine structure constant. These results are based on the Feynman diagrams [11-16,41] of Fig. 2 as will be described in more detail below. Roughly speaking, the gravitational potential changes the energy of a virtual electron-positron pair, which in turn produces a small change $\delta E(k)$ in the energy of a photon with wave vector $k$ as can be shown using perturbation theory. This results in a small correction to the angular frequency $\omega(k)$ of a photon and thus its velocity $c = \omega(k)/k$. The analogous effects for neutrinos involve the weak interaction and they are negligibly small in comparison. As a result, this model predicts a small but observable reduction in the velocity of photons relative to that of neutrinos. In principle, the reduction in the speed of light could be directly measured by a local observer, but a small change in $c$ can be more easily observed by comparing the photon and neutrino velocities.

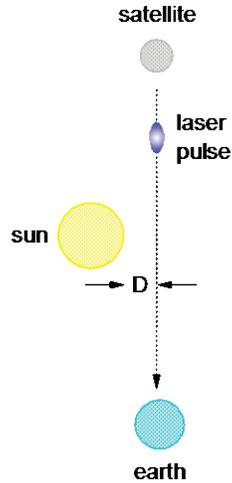

Fig. 1. A measurement of the transit time at the speed of light from a distant satellite to earth. Einstein predicted that the speed of light as measured in a global reference frame would be reduced by the gravitational potential of the sun as described by Eq. (1), which is in good agreement with experiments. Here $D$ is the distance of closest approach. The deflection of the light beam by the gravitational potential of the sun is very small and is not illustrated here.

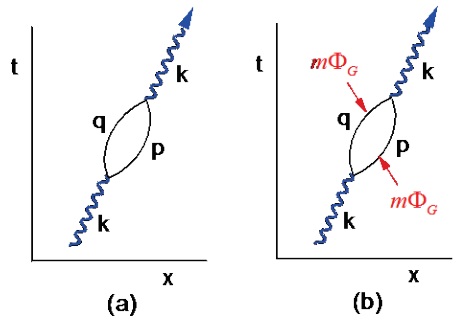

Fig. 2. (a) A Feynman diagram in which a photon with wave vector $\mathbf{k}$ is annihilated to produce a virtual state containing an electron with momentum $\mathbf{p}$ and a positron with momentum $\mathbf{q}$. After a short amount of time, the electron and positron are annihilated to produce a photon with the original wave vector $\mathbf{k}$. Any effect that this process may have on the velocity of light is removed using renormalization techniques to give the observed value of $c_0$. (b) The same process, except that now the energies of the virtual electron and positron include their gravitational potential energy $m\Phi_G$, as indicated by the arrows. This produces a small change in the velocity of light that is experimentally observable. The variable $t$ represents the time while $x$ represents the position in three dimensions (in arbitrary units).

The remainder of the paper begins with the motivation for including the gravitational energy $m\Phi_G$ in the Hamiltonian of a quantum system. The correction to the speed of light due to the gravitational potential is then calculated using quantum electrodynamics and standard perturbation theory in Section 3. Gauge invariance and the equivalence principle are discussed in Section 4, while the predicted delay in the photon arrival time for Supernova 1987a is calculated in Section 5 and compared with the experimental data. A summary and conclusions are presented in Section 6. Appendix A discusses the role of gravitational potentials in the theory of general relativity and the well-known analogy with electromagnetism. Appendix B shows that the relativistic Hamiltonian considered here correctly reduces to the Pauli equation in the nonrelativistic limit for both electrons and positrons.

## 2. Gravitational potentials in the quantum theory

In nonrelativistic quantum mechanics, the Hamiltonian $\hat{H}$ for a particle with mass $m$ in a Newtonian gravitational field is given by

$$\hat{H} = -\frac{\hbar^2}{2m}\nabla^2 + m\Phi_G(\mathbf{r}) \tag{2}$$

where $\Phi_G(\mathbf{r})$ is the Newtonian gravitational potential at position $\mathbf{r}$. Eq. (2) has been successfully used to analyze the results of neutron interferometer experiments in a gravitational field, for example [1].

Eq. (2) can be generalized to include the non-Newtonian gravitational effects of a rotating mass $M$ by making use of the well-known analogy [33,42-49] between electromagnetism and Einstein's field equations for a weak gravitational field. As outlined in Appendix A, it is possible to define a gravitational vector potential $\mathbf{A}_G(\mathbf{r},t)$ and a gravitational scalar potential $\Phi_G(\mathbf{r},t)$ that are determined by the metric. The motion of a classical particle is then described by the usual Lorentz force equation, which provides a convenient way to visualize general relativistic effects such as frame-dragging. This suggests that Eq. (2) can be generalized to

$$\hat{H} = \frac{1}{2m}\left(\frac{\hbar}{i}\nabla - \frac{m}{c}\mathbf{A}_G(\mathbf{r},t)\right)^2 + m\Phi_G(\mathbf{r},t). \tag{3}$$

Eq. (3) has previously been used in connection with the gravitational analog of the Aharonov-Bohm effect, for example [33,47,50-52]. We will be interested here in situations where the source of the gravitational field is stationary in the chosen coordinate frame, in which case $\mathbf{A}_G = 0$ and Eq. (3) reduces to Eq. (2).

Eqs. (2) and (3) suggest that it may be possible to represent the effects of a weak gravitational field in quantum electrodynamics, at least to a first approximation, by including $m\Phi_G$ in the usual interaction Hamiltonian $\hat{H}'$. This gives

$$\hat{H}' = -\frac{1}{c}\int d^3\mathbf{r}\,\hat{\mathbf{j}}_E(\mathbf{r},t)\cdot\hat{\mathbf{A}}_E(\mathbf{r},t) + \int d^3\hat{\mathbf{r}}\hat{\rho}_E(\mathbf{r},t)\hat{\Phi}_E(\mathbf{r},t) + \int d^3\hat{\mathbf{r}}\hat{\rho}_G(\mathbf{r},t)\Phi_G(\mathbf{r},t) \tag{4}$$

in the Lorentz gauge [12,16]. Here the electromagnetic charge density $\hat{\rho}_E(\mathbf{r},t)$ and current density $\hat{\mathbf{j}}_E(\mathbf{r},t)$ are given as usual by

$$\hat{\rho}_E(\mathbf{r},t) = q\hat{\psi}^\dagger(\mathbf{r},t)\hat{\psi}(\mathbf{r},t)$$
$$\hat{\mathbf{j}}_E(\mathbf{r},t) = cq\hat{\psi}^\dagger(\mathbf{r},t)\boldsymbol{\alpha}\hat{\psi}(\mathbf{r},t). \tag{5}$$

The charge of an electron is denoted by $q$, $\hat{\psi}(\mathbf{r},t)$ is the Dirac field operator, $\boldsymbol{\alpha}$ represents the Dirac matrices [53], and $\hat{\rho}_G(\mathbf{r},t)$ corresponds to the mass density of the particles as described in more detail in Appendix B. $\hat{\mathbf{A}}_E(\mathbf{r},t)$ and $\hat{\Phi}_E(\mathbf{r},t)$ represent the vector and scalar potentials of the electromagnetic field and the first two terms in Eq. (4) correspond to the usual interaction between charged particles and the electromagnetic field. The third term represents the gravitational potential energy of any particles. It is shown in Appendix B that the interaction Hamiltonian of Eq. (4) correctly reduces to the Pauli equation in the nonrelativistic limit with the correct sign of $m\Phi_G$ for both electrons and positrons. This is equivalent to the Schrodinger equation of Eq. (2) in the absence of a magnetic field.

For our purposes, the gravitational potentials $\mathbf{A}_G$ and $\Phi_G$ will be assumed to be classical fields. Similar results would be obtained if a weak gravitational field were quantized to introduce gravitons, as is illustrated in Fig. 3.

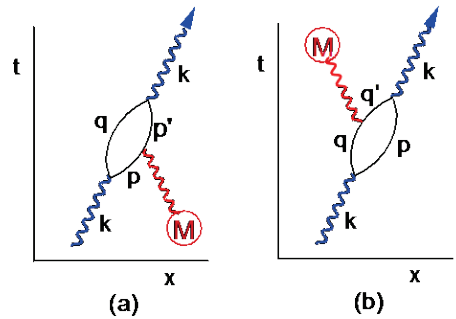

Fig. 3. Feynman diagrams that are equivalent to those of Fig. 2 except that here gravity is quantized and the gravitational potential is produced by the emission and absorption of virtual gravitons. (a) A graviton produced by mass $M$ is absorbed by a virtual electron, changing its momentum from $\mathbf{p}$ to $\mathbf{p'}$. (b) A virtual positron emits a graviton that is then absorbed by mass $M$. The momentum of the positron is changed from $\mathbf{q}$ to $\mathbf{q'}$. Two other diagrams (not shown) involve the emission of a graviton by a virtual electron or the absorption of a graviton by a virtual positron.

Eq. (4) represents a simple model in which the gravitational potential of any massive particles is included in the Hamiltonian. Although this assumption seems plausible, it leads to an observable correction to the speed of light that is not equivalent to what is obtained using the currently-accepted generalization of the Dirac equation to curved spacetime [17-19]. The fact that the predictions of this simple model are in reasonable agreement with experimental observation may provide some motivation for considering the differences between these two approaches.

## 3. Calculated correction to the speed of light

In quantum electrodynamics, there is a probability amplitude for a photon propagating in free space with wave vector **k** and angular frequency $\omega = ck$ to be annihilated while producing a virtual state containing an electron-positron pair, as illustrated in Fig. 2a. The virtual state only exists for a brief amount of time, after which the process is reversed and the electron-positron pair is annihilated and the original photon is reemitted. This process, which is known as vacuum polarization [54], leads to divergent terms that can be eliminated using renormalization techniques while small corrections to this process can produce observable effects.

Here we will calculate the change $\Delta E$ in the energy of a photon with wave vector **k** due to the interaction of a virtual electron-positron pair with the gravitational potential as illustrated schematically in Fig. 2b. The gravitational potential changes the energy of the virtual electron–positron state by $2m\Phi_G$, as is shown in Appendix B. That in turn changes the energy of the photon by a small amount as will be shown below using perturbation theory. The dependence of the energy (and thus the frequency $\omega$) of a photon on $k$ will produce a correction to its velocity $\omega / k$. It will be assumed that the gravitational field is not quantized and is described by the Newtonian potential $\Phi_G$.

From second-order perturbation theory [53], the change $\Delta E^{(2)}$ in the energy of a photon with wave vector **k** is given by

$$\Delta E^{(2)} = \sum_n \frac{\langle \mathbf{k} | \hat{H}' | n \rangle \langle n | \hat{H}' | \mathbf{k} \rangle}{E_0^{(0)} - E_n^{(0)}}. \tag{6}$$

Here $|\mathbf{k}\rangle$ represents the unperturbed initial state containing only the photon while $|n\rangle$ represents all possible intermediate states containing an electron-positron pair. The unperturbed energy of the initial state is represented by $E_0^{(0)}$ while $E_n^{(0)}$ is the unperturbed energy of the intermediate state. For the purposes of this calculation, the gravitational potential term in Eq. (4) will be included in the unperturbed Hamiltonian $\hat{H}_0$ while the electromagnetic interaction terms will be included in the perturbation Hamiltonian $\hat{H}'$. As a result, the unperturbed energy of the intermediate state containing an electron-positron pair becomes $E_n^{(0)} = E_p + E_q + 2m\Phi_G$. Here **p** and **q** are the momenta of the virtual electron and positron respectively, as in Fig. 2, while $E_\mathbf{p}$ is the relativistic energy of a free particle given as usual by $E_p = \sqrt{p^2 c^2 + m^2 c^4}$.

Straightforward perturbation theory will be used for simplicity and because the usual Feynman diagram rules [11-16] may not be directly applicable to the Hamiltonian of Eq. (4). Eq. (6) corresponds to steady state perturbation theory, but the same results can be obtained using the forward-scattering amplitude from time-dependent perturbation theory [53].

We will use periodic boundary conditions with a unit volume $V$ [55]. In the Schrodinger picture, the Dirac field operators are then given by [15]

$$\hat{\psi}(\mathbf{r}) = \sum_{\mathbf{p},s} \sqrt{\frac{mc^2}{E_{\mathbf{p}}}} \left[ \hat{b}(\mathbf{p},s) u(\mathbf{p},s) e^{i\mathbf{p}\cdot\mathbf{r}/\hbar} + \hat{c}^\dagger(\mathbf{p},s) v(\mathbf{p},s) e^{-i\mathbf{p}\cdot\mathbf{r}/\hbar} \right]. \tag{7}$$

Here $\hat{b}^\dagger(\mathbf{p},s)$ creates an electron with momentum $\mathbf{p}$ and spin $s$, whose values will be denoted by $\pm$ to indicate spin up or down, while $\hat{c}^\dagger(\mathbf{p},s)$ creates a positron with momentum $\mathbf{p}$ and spin $s$. The Dirac spinors $u(\mathbf{p},s)$ and $v(\mathbf{p},s)$ are defined [14,15,53] by

$$u(\mathbf{p},+) = \sqrt{\frac{E_{\mathbf{p}}+mc^2}{2mc^2}} \begin{bmatrix} 1 \\ 0 \\ \dfrac{p_z c}{E_{\mathbf{p}}+mc^2} \\ \dfrac{p_+ c}{E_{\mathbf{p}}+mc^2} \end{bmatrix} \qquad u(\mathbf{p},-) = \sqrt{\frac{E_{\mathbf{p}}+mc^2}{2mc^2}} \begin{bmatrix} 0 \\ 1 \\ \dfrac{p_- c}{E_{\mathbf{p}}+mc^2} \\ \dfrac{-p_z c}{E_{\mathbf{p}}+mc^2} \end{bmatrix}$$

$$v(\mathbf{p},-) = \sqrt{\frac{E_{\mathbf{p}}+mc^2}{2mc^2}} \begin{bmatrix} \dfrac{p_z c}{E_{\mathbf{p}}+mc^2} \\ \dfrac{p_+ c}{E_{\mathbf{p}}+mc^2} \\ 1 \\ 0 \end{bmatrix} \qquad v(\mathbf{p},+) = \sqrt{\frac{E_{\mathbf{p}}+mc^2}{2mc^2}} \begin{bmatrix} \dfrac{p_- c}{E_{\mathbf{p}}+mc^2} \\ \dfrac{-p_z c}{E_{\mathbf{p}}+mc^2} \\ 0 \\ 1 \end{bmatrix} \tag{8}$$

where $p_\pm \equiv p_x \pm i p_y$.

The scalar electromagnetic potential $\hat{\Phi}_E(\mathbf{r})$ and the longitudinal part of $\hat{\mathbf{A}}_E(\mathbf{r})$ do not contribute to this process. In Gaussian units, the transverse part of the electromagnetic vector potential operator is given by

$$\hat{\mathbf{A}}_E(\mathbf{r}) = \sum_{\mathbf{k},\boldsymbol{\varepsilon}} \sqrt{\frac{2\pi\hbar c^2}{\omega}} \left[ \hat{a}_{\mathbf{k}} \boldsymbol{\varepsilon} e^{i\mathbf{k}\cdot\mathbf{r}} + \hat{a}_{\mathbf{k}}^\dagger \boldsymbol{\varepsilon} e^{-i\mathbf{k}\cdot\mathbf{r}} \right]. \tag{9}$$

Here $\omega = ck$ and $\boldsymbol{\varepsilon}$ denotes two transverse polarization unit vectors. Without loss of generality, we can assume that the initial photon has its wave vector $\mathbf{k}$ in the $\hat{x}$ direction with its polarization along the $\hat{z}$ direction.

The integral over $\mathbf{r}$ in the interaction Hamiltonian of Eq. (4) combined with the exponential factors in $\hat{\psi}^\dagger(\mathbf{r})$, $\hat{\psi}(\mathbf{r})$, and $\hat{\mathbf{A}}_E(\mathbf{r},t)$ give a delta-function that conserves momentum, so that $\mathbf{q} = \hbar\mathbf{k} - \mathbf{p}$ and the sum over intermediate states reduces to a sum over all values of $\mathbf{p}$. We will assume that the energy of the photon is sufficiently small that $\hbar kc << mc^2$, in which case $E_\mathbf{q} \doteq E_\mathbf{p}$. (I.e., the recoil momentum from absorbing the photon has a negligible effect on the virtual particle energies) For the same reason, we can approximate $\mathbf{q}$ by $-\mathbf{p}$ in the evaluation of $u^\dagger(\mathbf{p},s)\alpha_z v(\mathbf{q},s')$ with the result that

$$u^\dagger(\mathbf{p},-)\alpha_z v(\mathbf{q},+) = -\left(\frac{E_p + mc^2}{2mc^2}\right)\left(1 + \frac{(p_x^2 + p_y^2 - p_z^2)c^2}{(E_p + mc^2)^2}\right), \tag{10}$$

with similar results for the other spin states. Here we have made use of Eq. (8) and the fact that

$$\alpha_z = \begin{bmatrix} 0 & 0 & 1 & 0 \\ 0 & 0 & 0 & -1 \\ 1 & 0 & 0 & 0 \\ 0 & -1 & 0 & 0 \end{bmatrix} \tag{11}$$

in the usual representation. Eq. (10) was derived by simply multiplying the relevant matrices and vectors, while the same results could have been obtained more generally by using the properties of the Dirac matrices. Combining these results with Eqs. (4), (7), and (9) gives

$$\langle n|H'|\mathbf{k}\rangle = \frac{q}{2}\sqrt{\frac{2\pi\hbar c^2}{\omega}}\left(\frac{E_p + mc^2}{E_p}\right)\left(1 + \frac{(p_x^2 + p_y^2 - p_z^2)c^2}{(E_p + mc^2)^2}\right) \tag{12}$$

for the spin combination of Eq. (10).

Inserting Eq. (12) into Eq. (6) and summing over all of the intermediate spin states gives the correction to the photon energy as

$$\Delta E^{(2)} = \frac{1}{(2\pi)^3}\frac{\alpha\pi c^3}{\hbar\omega_0}\int_{-\infty}^{\infty} d^3\mathbf{p}\,\frac{1}{\left(\hbar\omega_0 - 2E_p - 2m\Phi_G(\mathbf{r}_0)\right)}$$
$$\times \frac{(E_p + mc^2)^2}{E_p^2}\frac{\left[(E_p + mc^2)^4 + \frac{2}{3}(E_p + mc^2)^2 p^2c^2 + p^4c^4\right]}{(E_p + mc^2)^4} \tag{13}$$

Here we have introduced the fine structure constant $\alpha \equiv q^2/\hbar c$ and the factor of $1/(2\pi)^3$ comes from converting the sum to an integral. The notation $\omega_0$ has been used here to indicate that it is the unperturbed photon energy $\hbar\omega_0$ that appears in Eq. (13).

We can now use the assumption that $\hbar\omega_0 << E_p$ and $|m\Phi_G| << E_p$ to expand the denominator in the first term inside the integral of Eq. (13) in a Taylor series to first order in $m\Phi_G$ and to second order in $\hbar\omega_0$. This gives

$$\frac{1}{(\hbar\omega_0 - 2E_p - 2m\Phi_G)} = \frac{1}{E_p}\frac{m\Phi_G}{E_p}\left[\frac{1}{2} + \frac{1}{2}\frac{\hbar\omega_0}{E_p} + \frac{3}{8}\left(\frac{\hbar\omega_0}{E_p}\right)^2 + ...\right]. \tag{14}$$

We have only retained terms proportional to $m\Phi_G$ in Eq. (14), since we are only interested in the first-order effects of the gravitational field.

We will first consider the effects of the last term in Eq. (14) and then return to consider the remaining two terms. The contribution from the last term gives

$$\Delta E^{(2)} = \frac{3}{16\pi}\alpha c(\hbar\omega_0)(m\Phi_G)\int_0^\infty dp \frac{p^2c^2}{E_p^4}\frac{(E_p + mc^2)^2}{E_p^2}$$
$$\times \frac{\left[(E_p + mc^2)^4 + \frac{2}{3}(E_p + mc^2)^2 p^2c^2 + p^4c^4\right]}{(E_p + mc^2)^4}. \tag{15}$$

Evaluating the integral gives

$$\Delta E^{(2)} = \frac{9}{64}\alpha(\hbar\omega_0)\frac{\Phi_G}{c_0^2}. \tag{16}$$

The velocity of light is given by $c = \omega(k)/k$ and the correction $\Delta c$ to $c$ is thus

$$\Delta c = \frac{\Delta\omega(k)}{k} = \frac{\Delta E^{(2)}/\hbar}{\omega_0/c_0}. \tag{17}$$

Inserting the value of $\Delta E^{(2)}$ from Eq. (16) into Eq. (17) gives

$$\frac{\Delta c}{c_0} = \frac{9}{64}\alpha\frac{\Phi_G}{c^2}. \tag{18}$$

Eq. (18) is the main result of this paper. Since $\Phi_G$ is negative, this gives a small reduction in the speed of light.

Returning to the first and second terms on the right-hand side of Eq. (14), it can be shown that their contributions to $\Delta c$ are proportional to $1/\omega_0^2$ and $1/\omega_0$, respectively. These are nonphysical terms that become infinite in the limit of long wavelengths, which is somewhat similar to the usual infrared divergences encountered in quantum electrodynamics. We can make an intuitive argument that these terms should vanish as a result of renormalization as follows: The loop diagram of Fig. 2a would give an infinite correction to the energy of a photon, so that the "bare" energy $\hbar\omega_B$ of the photon must be infinitely large as well. Identifying $\hbar\omega_0$ with $\hbar\omega_B$ would therefore cause the nonphysical terms that involve $1/\omega_B^2$ and $1/\omega_B$ to vanish, whereas $\hbar\omega_B$ cancels out of the finite correction of Eq. (17). A more rigorous treatment of renormalization would clearly be desirable, but that may not be possible in view of the fact that quantum gravity appears to be nonrenormalizable.

A neutrino can also undergo a virtual process in which particles such as W bosons, Z bosons, and leptons are created, after which the virtual particles are annihilated to give back the original neutrino state. The energy of the particles in the intermediate state will include their gravitational potential energy $m\Phi_G$, which will produce a small correction to the velocity of a neutrino that is analogous to that of a photon calculated above. But this process involves the weak interaction where the matrix elements are many orders of magnitude smaller than those for the electromagnetic interaction responsible for virtual electron-positron pair production. As a result, the expected correction to the velocity of neutrinos is negligible compared to that of photons.

## 4. Gauge invariance and the equivalence principle

Before we consider the magnitude of this effect, it is important to note that the results of this calculation are not gauge invariant with respect to the gravitational field. Conventional quantum electrodynamics is gauge invariant only because charge is conserved via

$$\nabla \cdot \hat{\mathbf{j}}_E = -\frac{\partial \hat{\rho}_E}{\partial t}. \tag{19}$$

As a result, creating an electron-positron pair in a region of uniform electrostatic potential has no effect on the total electrostatic energy of the system because there is no change in the total charge, as illustrated in Fig. 4a. But the Hamiltonian of quantum electrodynamics does not conserve the mass of the system in a virtual state containing an electron-positron pair and there is no equivalent of Eq. (19) for mass in that case, as is discussed in more detail in Appendix B. Thus the creation of an electron-positron pair in a uniform gravitational potential does change the energy of the system if we assume the Hamiltonian of Eq. (4), as illustrated in Fig. 4b. This explains why the predicted change in the velocity of light in Eq. (16) depends on the value of the gravitational potential and not just the gravitational field in violation of gauge invariance.

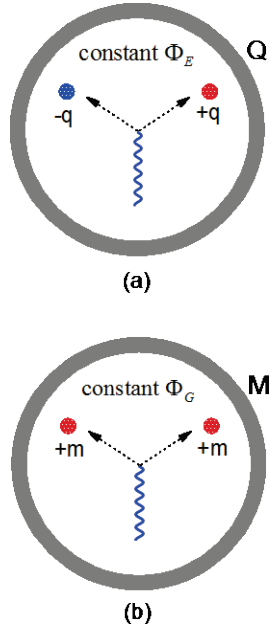

**(a)**

**(b)**

Fig. 4. Effects of a constant electrostatic or gravitational potential on the energy of the electron-positron pair produced in the Feynman diagrams of Figs. 2 and 3. (a) A region of constant electrostatic potential $\Phi_E$ is created using a uniform spherical charge distribution with a total charge of $Q$. The energy of an electron-positron pair is unaffected by $\Phi_E$ because the total change in the charge is zero as required by Eq. (19). (b) A region of constant gravitational potential $\Phi_G$ is created using a uniform spherical mass distribution with a total mass of $M$. Now the energy of the electron-positron pair is changed by $2m\Phi_G$ because the pair production process does not conserve mass. There is no equivalent of Eq. (19) in this case.

Based on the equivalence principle [38,39,56], one would expect that these effects should vanish in a local freely-falling reference frame. The calculations described above were performed using a coordinate frame that was assumed to be at rest with respect to mass $M$, where it seems reasonable to suppose that the effects of gravity can be represented by the Feynman diagrams of Figs. 2b or Fig 3. In that reference frame, the photons and neutrinos would travel at different velocities according to Eq. (18). If we made a transformation to a local freely-falling coordinate frame where the laws of physics are assumed to be the same as in the absence of a gravitational field, then the photons and neutrinos would be expected to travel at the same velocity. This leads to a contradiction, since there can be no disagreement as to whether or not two particles are travelling at the same velocity. Thus the Hamiltonian of Eq. (4) leads to a small departure from the equivalence principle, which is closely related to the lack of gravitational gauge invariance noted above.

It has been predicted [57-58] that electrons with spin up and spin down will fall at different rates in a gravitational field in apparent violation of the weak equivalence principle. (This effect is analogous to the spin-orbit coupling of an electron moving in the Coulomb field of an atom.) That may not be too surprising given that a classical object with nonzero angular momentum will exhibit similar effects [58,59]. But the fact that an electron is a point particle makes this situation different from that of a gyroscope whose finite extent makes it susceptible to tidal forces, for example. The weak equivalence principle is sometimes stated as only applying to point particles with zero angular momentum, in which case it should not be applied to electrons. This raises some questions regarding the assumptions that are inherent in the derivation of the generally covariant form of the Dirac equation. In any event, the predicted correction to the speed of light from Eq. (18) provides further motivation for experimental tests of the equivalence principle.

## 5. Comparison with experimental observations

The first neutrinos from Supernova 1987a arrived 7.7 hours before the first photons. The currently-accepted interpretation [26] of this data is that the first burst of neutrinos must not have been associated with the supernova because there is no conventional explanation for how the neutrinos could have arrived at that time. If Eq. (18) is valid, it could explain this long-standing anomaly.

The value of $\Phi_G / c^2$ is needed in order to compare the predicted correction to the speed of light with experimental observations such as those from Supernova 1987a. The Newtonian gravitational potential from an object with mass $M$ at a distance $R$ is given by

$$\frac{\Phi_G}{c^2} = -\frac{GM}{Rc^2} \qquad (20)$$

where $G$ is the gravitational constant. Table 1 shows the approximate value of $\Phi_G / c^2$ and the corresponding correction to the speed of light from Eq. (18) for the case in which the source of the gravitational potential is the earth, the sun, or the Milky Way galaxy. It can be seen that the contributions to the gravitational potential from the earth and sun are negligible compared to that of the Milky Way galaxy.

The gravitational potential $\Phi_{GU}$ from the universe as a whole is not given by the Newtonian formula of Eq. (20). Instead, $\Phi_{GU} = 0$ for a flat universe as is discussed in Appendix A. Astronomical observations indicate that the universe is flat to within the experimental uncertainty, in which case the only contribution to the gravitational potential is from local variations in the mass density, such as the Milky Way galaxy. Mass variations at larger distances appear to be negligible in comparison.

|        | $\Phi_G / c^2$          | $\Delta c / c_0$        |
|--------|-------------------------|-------------------------|
| Earth  | $-6.4 \times 10^{-10}$  | $-6.6 \times 10^{-13}$  |
| Sun    | $-9.9 \times 10^{-9}$   | $-1.01 \times 10^{-11}$ |
| Galaxy | $-4.2 \times 10^{-6}$   | $-4.3 \times 10^{-9}$   |

Table 1. Gravitational potential $\Phi_G / c^2$ and fractional correction to the speed of light $\Delta c / c_0$ from the earth, the sun, and the Milky Way galaxy. The value of the gravitational potential from the Milky Way galaxy was approximated at the location of the earth using Eq. (20).

Supernova 1987a was located in the Large Magellanic Cloud [60], which is a smaller galaxy that is gravitationally bound to the Milky Way galaxy. In order to predict the expected difference in the arrival times of photons and neutrinos at the earth, it is necessary to integrate the effects of Eq. (18) over their path which is illustrated in Fig. 5. Longo [61] integrated the usual relativistic factor of $2\Phi_G(\mathbf{r}) / c_0^2$ in Eq. (1) over the path illustrated in Fig. 5 using a model for the gravitational potential produced by the Milky Way galaxy. (The contribution of the Large Magellanic Cloud to the gravitational potential is negligible due to its small mass.) He obtained a total time delay of 3506 hours from the usual correction to the speed of light in Eq. (1). A similar calculation by Krauss and Tremaine [62] gave a total time delay of 3944 hours.

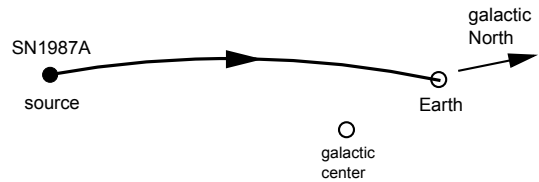

Fig. 5. Path followed by the neutrinos and photons from Supernova 1987a, which was located in the Large Magellanic Cloud [62]. Refs. [61] and [62] estimated the time delay expected from Eq. (1) using the gravitational potential from the Milky Way galaxy, which can then be used to calculate the contribution from Eq. (18).

The integral of Eq. (18) over the same path differs from these estimates by a factor of $9\alpha / 64$ and also by a factor of $1/2$, since the factor of 2 in Eq. (1) does not appear in Eq. (18). Applying this factor to the average of the results of Longo [61] and of Krauss and Tremaine [62] gives a predicted delay of 1.9 hours for the photons relative to the neutrinos based on Eq. (18). This estimate is really a lower bound on the actual delay, since Refs. [61] and [62] only included the mass of the Milky Way that is within 60 kpc of the center of the galaxy. That represents roughly half of the estimated mass of the galaxy and the predicted delay could be as large as 4 hours if the additional mass were included. (The effects of dark matter appear to be included in Refs. [61] and [62] and no correction for that is required.)

The observations made during Supernova 1987a are illustrated in Fig. 6, which is based on a review article by Bahcall and his colleagues [26]. A burst of neutrinos was observed by a detector underneath Mont Blanc followed 4.7 hours later by a second burst of neutrinos that was detected in the Kamiokande II detector in Japan and the IMB detector in Ohio. The first observation of visible light from the supernova was then observed approximately three hours after the second burst of neutrinos, or 7.7 hours after the first burst of neutrinos. As mentioned earlier, the usual interpretation of this data is that the first burst of neutrinos must not have been associated with the supernova for the reasons described below.

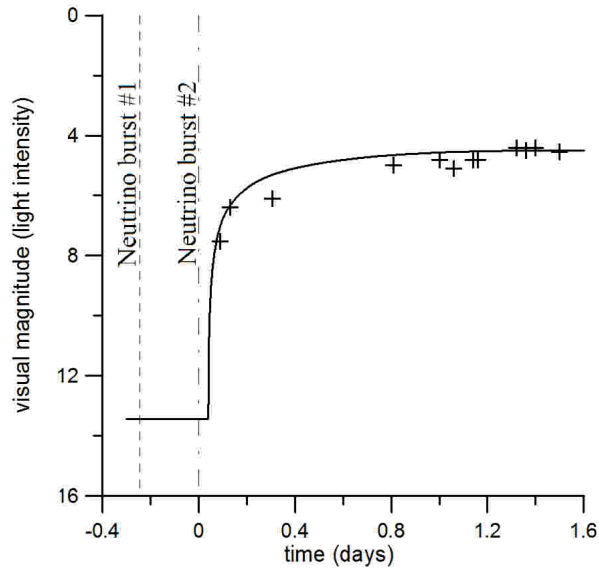

Fig. 6. Sequence of events observed during Supernova 1987a [26]. The time at which the first burst of neutrinos was observed in the Mont Blanc detector is indicated by the dashed line, while the time at which the second burst of neutrinos was observed in the Kamiokande II and IMB detectors is indicated by the dotted-dashed line. The data points show the magnitude (logarithmic intensity) of the observed visible light from the supernova as a function of the time (in days) after the arrival of the second burst of neutrinos. The solid line is the result of a numerical calculation based on the accepted model of the supernova. The first burst of neutrinos was considered to be inconsistent with the accepted model and was rejected as a statistical outlier [26]. This discrepancy could be explained if the arrival of the photons was delayed as predicted by Eq. (18).

A numerical simulation of the collapse of the progenitor star gave a predicted visible light intensity as a function of time (light curve) that is represented by the solid line in Fig. 6. There is an expected time delay of approximately three hours between the collapse of the core and the production of visible light at the surface of the star due to the propagation of a shock wave through the stellar material. (Any light produced in the interior of the star will be prevented from immediately reaching the surface due to diffusion.) As a result, Bahcall and his colleagues have stated that the arrival time of the first burst of neutrinos "is not consistent with the observed light curve" [26]. In addition, the fact that the first burst of neutrinos was only detected by the Mont Blanc detector and not the other two detectors, which were assumed at the time to have higher sensitivities, further suggested that the first burst of neutrinos must have been an anomaly that was not associated with Supernova 1987a [26].

The probability that the detection of the initial burst of neutrinos in the Mont Blanc detector was a random occurrence has been estimated to be less than $10^{-4}$ [27]. As a result, there are some experts in the field who consider the origin of the first burst of neutrinos to be an open question [27,63-65]. A more recent numerical simulation [65] showed that a progenitor star with a sufficiently high rate of angular rotation would be expected to produce an initial incomplete collapse of the core followed by a second collapse, which would produce two bursts of neutrinos instead of just one. In addition, the simulation showed that different kinds of neutrinos with different energy ranges should have been produced during the two collapses [64,65]. The material used in the Mont Blanc detector was different from that used in the other two detectors and the expected sensitivity of detection for the kind of neutrinos in the first burst has been estimated to be a factor of 20 higher in the Mont Blanc detector than the other detectors, which is consistent with the observations [64].

The possibility of a double collapse of the core suggests an alternative explanation for the observations associated with Supernova 1987a. In this scenario, the first burst of neutrinos signaled the initial collapse of the core with an associated production of visible light roughly 3 hours later as expected from the models. If the photons were delayed by an additional 4.7 hours by the gravitational potential in Eq. (18), then the light would have arrived 7.7 hours after the first neutrino burst, as observed. The second collapse of the core would have produced an increase in the intensity of the visible light approximately 4.7 hours after the arrival of the first photons. This is consistent with the observation that the light signal increased more rapidly than would have otherwise been expected during that time interval [26].

The photon delay of 1.9 hours relative to the neutrinos as predicted by Eq. (18) is only 40% of the 4.7 hour delay assumed in the scenario described above. As mentioned earlier, this estimate is a lower bound on the actual delay which could be as large as 4 hours. Thus Eq. (18) is in reasonable agreement with the experimental observations and it provides a possible explanation for the first burst of neutrinos that is inconsistent with the conventional model of the supernova.

The Hamiltonian of Eq. (4) does not appear to be ruled out by the results of existing high-precision tests of quantum electrodynamics [66]. It would result in a small correction to the anomalous magnetic moment of the electron, for example, that is much smaller than the precision of the current experiments [28-29]. The model would also predict [66] a correction to the decay rate of orthopositronium that is approximately two orders of magnitude smaller than the current experimental precision [67]. Future experiments of that kind may eventually allow an independent test of the implications of including the gravitational potential in the Hamiltonian.

## 6. Summary and conclusions

A simple model has been considered here in which it was assumed that, to a first approximation, the effects of a weak gravitational field on a quantum system can be represented by including the gravitational potential energy of any massive particles in the Hamiltonian.

When applied to the Hamiltonian of quantum electrodynamics, this results in virtual electrons and positrons having a gravitational potential energy that is the same as that of a real particle. Perturbation theory was then used to show that such a model predicts a small reduction in the speed of light while the corresponding effects for neutrinos are negligibly small due to their weak interactions.

The predicted correction to the speed of light depends on the gravitational potential and not the gravitational field. An observable difference between the velocity of photons and neutrinos that depends only on the gravitational potential is not gauge invariant. The origin of this lack of gauge invariance can be understood from the fact that the gravitational potential energy has the same sign for the virtual electrons and positrons created during pair production, while their electrostatic potential energies have the opposite sign and cancel out, as illustrated in Fig. 4.

The predictions of this simple model are also in disagreement with the analogous calculations performed using the generalization of the Dirac equation to curved spacetime, which gives a much smaller effect that does depend on the gravitational field and not the potential itself [17-19]. The lack of gauge invariance and the disagreement with the generally-covariant form of the Dirac equation both suggest that this simple model must be nonphysical.

Nevertheless, the predictions of this model are in reasonable agreement with the experimental observations from Supernova 1987a, in which the first neutrinos arrived 7.7 hours before the first photons. There is no conventional explanation for how that could have occurred and the currently-accepted interpretation is that the first burst of neutrinos must not have been related to the supernova [26], despite the fact that the probability of such an event occurring at random has been estimated to be less than $10^{-4}$ [27]. The correction to the speed of light from Eq. (18), if correct, would explain this anomaly.

The differences between this simple model and the generally covariant form of the Dirac equation are closely related to the role of the equivalence principle, since photons and neutrinos should travel at the same velocity in a local freely-falling reference frame and thus in all reference frames. (The rest mass of a high-energy neutrino is negligible in this regard.) There is already considerable interest in experimental tests of the equivalence principle and the results of the model considered here may provide further motivation for experiments of that kind.

Quantum mechanics and general relativity are two of the most fundamental laws of physics. Combining these two theories in a consistent way is currently one of the major goals of physics research. The predicted correction to the speed of light in a gravitational potential may be of further interest if the currently-accepted principles of quantum mechanics and general relativity are eventually found to be incompatible in some way.

**Acknowledgements**

I would like to acknowledge stimulating discussions with W. Cohick, A. Kogut, and D. Shortle.

**Appendix A.  Gravitational potentials in general relativity**

This appendix briefly reviews the analogy between general relativity and electromagnetism for a weak gravitational field, which leads to the introduction of the gravitational analogs of the electromagnetic vector and scalar potentials. The field equations of general relativity are nonlinear but they can be linearized if the gravitational field is sufficiently small [33,38,39,42-49]. In that case, we can write the metric tensor $g_{\mu\nu}$ in the form

$$g_{\mu\nu} = \eta_{\mu\nu} + h_{\mu\nu} \tag{A1}$$

where $\eta_{\mu\nu}$ is the diagonal metric of special relativity with elements of $\pm 1$ and $h_{\mu\nu}$ is assumed to be small. It will also be convenient to define $h$ and $\overline{h}_{\mu\nu}$ by

$$h \equiv \eta^{\mu\nu} h_{\mu\nu}$$
$$\overline{h}_{\mu\nu} \equiv h_{\mu\nu} - \frac{1}{2}\eta_{\mu\nu}h. \tag{A2}$$

We can then define [43-45,49] the gravitational vector and scalar potentials $\mathbf{A}_G$ and $\Phi_G$ by

$$\Phi_G = -\frac{1}{4}c^2 \overline{h}_{00}$$
$$A_{Gi} = \frac{1}{4}c^2 \overline{h}_{0i}. \tag{A3}$$

We can also define two vectors $\mathbf{E}_G$ and $\mathbf{B}_G$ by

$$\mathbf{E}_G = -\nabla\Phi_G - \frac{1}{c}\frac{\partial}{\partial t}\mathbf{A}_G$$
$$\mathbf{B}_G = \nabla \times \mathbf{A}_G. \tag{A4}$$

Einstein's field equations can then be used to show that

$$\nabla \cdot \mathbf{E}_G = -4\pi G \rho_G$$
$$\nabla \cdot \mathbf{B}_G = 0$$
$$\nabla \times \mathbf{E}_G = -\frac{1}{c}\frac{\partial \mathbf{B}_G}{\partial t}$$
$$\nabla \times \mathbf{B}_G = \frac{1}{c}\frac{\partial \mathbf{E}_G}{\partial t} - \frac{4\pi G}{c}\mathbf{j}_G, \tag{A5}$$

where $\rho_G$ and $\mathbf{j}_G$ are the mass density and current. The potentials can be also be shown to obey the wave equations

$$\nabla^2 \Phi_G - \frac{1}{c^2}\frac{\partial^2 \Phi_G}{\partial t^2} = 4\pi G \rho_G$$

$$\nabla^2 \mathbf{A}_G - \frac{1}{c^2}\frac{\partial^2 \mathbf{A}_G}{\partial t^2} = \frac{4\pi G}{c}\mathbf{j}_G.$$

(A6)

Equations (A4) through (A6) are the same as those of classical electromagnetism except for the factor of $G$ and the sign of the source terms.

The geodesic equation can also be used to show that the trajectory of a particle of mass $m$ is given [43-45,49] by

$$m\frac{d^2\mathbf{r}}{dt^2} = \mathbf{f}_G = m\mathbf{E}_G + 4m\frac{1}{c}\frac{d\mathbf{r}}{dt}\times\mathbf{B}_G$$

(A7)

in the limit of low velocities. Here $\mathbf{f}_G$ is the gravitational force and Eq. (A7) is the same as the Lorentz force in electromagnetism except for the factor of 4. Some authors redefine a new vector potential $\mathbf{A'}_G = \mathbf{A}_G / 4$ and a new gravitomagnetic field $\mathbf{B'}_G = \mathbf{B}_G / 4$ in order to put Eq. (A7) into the same form as in electromagnetism. In that case the wave equation of Eq. (A6) no longer holds in its present form. This factor of 4 is due to the fact that the gravitational field is a tensor and not a vector, and its appearance somewhere in the equations is unavoidable.

One of the most interesting and fundamental features of the quantum theory is the fact that it is the electromagnetic potentials $\mathbf{A}_E$ and $\Phi_E$, not the electromagnetic fields $\mathbf{E}_E$ and $\mathbf{B}_E$, that appear in the Hamiltonian for a charged particle [53]. This gives rise to the Aharonov-Bohm effect [50] in which there are observable phenomena that occur in regions of space where $\mathbf{E}_E = 0$ and $\mathbf{B}_E = 0$, for example. This and the Lorentz force equation (A7) both suggest that the Hamiltonian for a nonrelativistic particle in a weak gravitational field should be taken to be

$$\hat{H} = \frac{1}{2m}\left(\frac{\hbar}{i}\nabla - \frac{4}{c}m\mathbf{A}_G\right)^2 + m\Phi_G$$

(A8)

in analogy with electromagnetism. Eq. (3) in the text results from replacing the vector potential with $\mathbf{A'}_G = \mathbf{A}_G / 4$. Eq. (A8) has previously been used in connection with the gravitational analog of the Aharonov-Bohm effect [33,47,50-52], for example.

It can be seen from Eq. (A3) that the gravitational potential $\Phi_{GU}$ from the universe as a whole is zero for a flat universe where $h_{\mu\nu} = 0$ aside from the effects of local mass density variations. This justifies the assumption in the main text that the total mass of the universe does not contribute to the correction to the speed of light as might be expected from the Newtonian expression of Eq. (20).

**Appendix B. Mass density operator and the nonrelativistic limit**

It was assumed in the text that the gravitational potential energy of a virtual electron-positron pair is $2m\Phi_G$ while the electrostatic potential energy cancels to zero. This difference between $\hat{\rho}_G(\mathbf{r})$ and $\hat{\rho}_E(\mathbf{r})$ is responsible for the predicted correction to the speed of light as well as the lack of gauge invariance in Fig. 4. The purpose of this Appendix is to define a mass density operator $\hat{\rho}_G(\mathbf{r})$ with these properties. The nonrelativistic limit of the interaction Hamiltonian of Eq. (4) is then shown to reduce to the Pauli equation with the correct sign of the gravitational potential energy for both electrons and positrons.

The electric charge and current densities can be written in covariant form as a four-vector $\hat{j}_E^{\ \mu}(\mathbf{r}) = qc\tilde{\psi}(\mathbf{r})\gamma^\mu\hat{\psi}(\mathbf{r})$ , whose fourth component is c times the charge density $\hat{\rho}_E(\mathbf{r}) = q\tilde{\psi}(\mathbf{r})\beta\hat{\psi}(\mathbf{r})$ . Here the $\gamma^\mu$ are the usual Dirac matrices and $\tilde{\psi}(\mathbf{r}) \equiv \hat{\psi}^\dagger(\mathbf{r})\beta$ is the adjoint field operator, where $\beta$ is given as usual by

$$\beta = \begin{bmatrix} 1 & 0 & 0 & 0 \\ 0 & 1 & 0 & 0 \\ 0 & 0 & -1 & 0 \\ 0 & 0 & 0 & -1 \end{bmatrix}. \tag{B1}$$

Roughly speaking, the minus signs in $\beta$ ensure that a positron will have the opposite charge from an electron. Since we want the gravitational potential to have the same sign for both electrons and positrons, this suggests that the simplest choice for $\hat{\rho}_G(\mathbf{r})$ may be

$$\hat{\rho}_G(\mathbf{r}) = m\tilde{\psi}(\mathbf{r})\hat{\psi}(\mathbf{r}). \tag{B2}$$

This differs from $\hat{\rho}_E(\mathbf{r})$ by the absence of $\beta$ which would be expected to give a positive gravitational potential for both electrons and positrons.

We will first investigate the nonrelativistic limit of the interaction Hamiltonian in Eq. (4). Combining $\hat{H}'$ with the remaining non-interacting terms in the Hamiltonian and inserting $\hat{\rho}_G(\mathbf{r})$ from Eq. (B2) gives the total Hamiltonian $\hat{H}$ :

$$\hat{H} = \int d^3r\, \hat{\psi}^\dagger \left[ c\boldsymbol{\alpha}\cdot\left(\frac{\hbar}{i}\nabla - \frac{q}{c}\mathbf{A}\right)\hat{\psi} + \beta\left(mc^2 + m\Phi_G\right)\hat{\psi} + q\Phi_E\hat{\psi} \right] + \sum_{\mathbf{k},\varepsilon}\left(\hat{a}_{\mathbf{k},\varepsilon}^\dagger\hat{a}_{\mathbf{k},\varepsilon} + 1/2\right)\hbar\omega_{\mathbf{k}}. \tag{B3}$$

This is the standard Hamiltonian of quantum electrodynamics [68] aside from the $m\Phi_G$ term. The time dependence of $\hat{\psi}(\mathbf{r},t)$ in the Heisenberg picture can be calculated using the anti-commutation property $\{\hat{\psi}_\alpha^\dagger(\mathbf{r},t),\hat{\psi}_\beta(\mathbf{r},t)\} = \delta_{\alpha\beta}\delta^3(\mathbf{r}-\mathbf{r}')$ , which gives [68]

$$i\hbar \frac{\partial \hat{\psi}}{\partial t} = \frac{1}{i\hbar}\left[\hat{\psi}, \hat{H}\right] = c\boldsymbol{\alpha} \cdot \left(\frac{\hbar}{i}\nabla - \frac{q}{c}\mathbf{A}\right)\hat{\psi} + \beta\left(mc^2 + m\Phi_G\right)\hat{\psi} + q\Phi_E\hat{\psi}. \qquad \text{(B4)}$$

This is the usual Dirac equation for the second-quantized field operator with the addition of the gravitational potential term.

The nonrelativistic limit of Eq. (B4) can now be calculated using the approach described in Ref. [53], for example. We first rewrite the four-component field operator in the form

$$\hat{\psi}(\mathbf{r},t) = \begin{bmatrix} \hat{\varphi}(\mathbf{r},t) \\ \hat{\chi}(\mathbf{r},t) \end{bmatrix} \qquad \text{(B5)}$$

where $\hat{\varphi}(\mathbf{r},t)$ and $\hat{\chi}(\mathbf{r},t)$ are two-component spinors. The Dirac equation (B4) is then equivalent to

$$i\hbar \frac{\partial \hat{\varphi}}{\partial t} = c\left(\frac{\hbar}{i}\nabla - \frac{q}{c}\mathbf{A}\right) \cdot \boldsymbol{\sigma}\hat{\chi} + \left(q\Phi_E + mc^2 + m\Phi_G\right)\hat{\varphi}$$
$$i\hbar \frac{\partial \hat{\chi}}{\partial t} = c\left(\frac{\hbar}{i}\nabla - \frac{q}{c}\mathbf{A}\right) \cdot \boldsymbol{\sigma}\hat{\varphi} + \left(q\Phi_E - mc^2 - m\Phi_G\right)\hat{\chi}, \qquad \text{(B6)}$$

where $\boldsymbol{\sigma}$ denotes the Pauli spin matrices. If we consider a positive-energy eigenstate corresponding to an electron with a velocity $v \ll c$, then $\hat{\chi} \ll \hat{\varphi}$ when acting on that state and the second line of Eq. (B5) gives to lowest order

$$\hat{\chi} = \frac{1}{2mc}\left(\frac{\hbar}{i}\nabla - \frac{q}{c}\mathbf{A}\right) \cdot \boldsymbol{\sigma}\hat{\varphi}. \qquad \text{(B7)}$$

Here we have assumed that the potential energies are small compared to the rest mass and that the time rate of change of the state is $-imc^2/\hbar$ to lowest order [53]. Inserting Eq. (B7) into the first line of Eq. (B6) and using the identity

$$\left(\mathbf{a}\cdot\boldsymbol{\sigma}\right)\left(\mathbf{b}\cdot\boldsymbol{\sigma}\right) = \left(\mathbf{a}\cdot\mathbf{b}\right) + i\boldsymbol{\sigma}\cdot\left(\mathbf{a}\times\mathbf{b}\right) \qquad \text{(B8)}$$

gives

$$i\hbar \frac{\partial \hat{\varphi}}{\partial t} = \frac{1}{2m}\left(\frac{\hbar}{i}\nabla - \frac{q}{c}\mathbf{A}\right)^2 \hat{\varphi} - \frac{q\hbar}{2mc}\boldsymbol{\sigma}\cdot\mathbf{B}\hat{\varphi} + \left(q\Phi_E + mc^2 + m\Phi_G\right)\hat{\varphi}. \qquad \text{(B9)}$$

This is the usual Pauli equation written in terms of (nonrelativistic) second-quantized field operators [53] with the addition of the gravitational potential term.

If we consider an eigenstate corresponding to a positron instead, then $\hat{\phi} << \hat{\chi}$ when acting on that state and the first line of Eq. (B6) gives to lowest order

$$\hat{\phi} = -\frac{1}{2mc}\left(\frac{\hbar}{i}\nabla - \frac{q}{c}\mathbf{A}\right)\cdot\boldsymbol{\sigma}\hat{\chi}. \tag{B10}$$

Inserting this into the second line of Eq. (B6) gives

$$i\hbar\frac{\partial\hat{\chi}}{\partial t} = -\frac{1}{2m}\left(\frac{\hbar}{i}\nabla - \frac{q}{c}\mathbf{A}\right)^2\hat{\chi} + \frac{q\hbar}{2mc}\boldsymbol{\sigma}\cdot\mathbf{B}\hat{\chi} + \left(q\Phi_E - mc^2 - m\Phi_G\right)\hat{\chi}. \tag{B11}$$

This can be rewritten by defining a new operator $\hat{\chi}' = \hat{\chi}^{\dagger}$ (charge conjugation) and taking the adjoint of Eq. (B11), which gives

$$i\hbar\frac{\partial\hat{\chi}'}{\partial t} = \frac{1}{2m}\left(\frac{\hbar}{i}\nabla + \frac{q}{c}\mathbf{A}\right)^2\hat{\chi}' - \frac{q\hbar}{2mc}\boldsymbol{\sigma}\cdot\mathbf{B}\hat{\chi}' + \left(-q\Phi_E + mc^2 + m\Phi_G\right)\hat{\chi}'. \tag{B12}$$

Eq. (B12) corresponds to the Pauli equation for a particle (a positron) whose charge and spin are opposite that of an electron but whose mass and gravitational potential energy are the same as that of an electron. Taking $\mathbf{A} = \Phi_E = 0$ gives the nonrelativistic Schrodinger equation of Eq. (2) as desired. Eqs. (B9) and (B12) show that this choice of $\hat{\rho}_G(\mathbf{r})$ gives the correct sign of the gravitational potential energy for both electrons and positrons at least in the nonrelativistic limit.

We now generalize this to the case in which the virtual electron-positron pair may have relativistic velocities. Here we make use of the fact that, in the text, the gravitational potential was included in the unperturbed Hamiltonian $\hat{H}_0$, where $\hat{H} = \hat{H}_0 + \hat{H}'$. The perturbation calculations were then based on the eigenstates of $\hat{H}_0$ and their corresponding energy eigenvalues $E_0$, which were assumed to include a gravitational potential energy of $m\Phi_G$ for each particle. First consider the value of the eigenvalue $E_0$ for a positron state $|\Psi\rangle = \hat{c}_{\mathbf{k}s}^{\dagger}|0\rangle$ with momentum $\hbar\mathbf{k}$ and spin $s$, where $|0\rangle$ is the vacuum. For a weak field with $\Phi_G / c^2 \ll 1$, the gravitational contribution $\Delta E_G$ to the eigenvalue $E_0$ is given [53] to lowest order in perturbation theory in $m\Phi_G$ by

$$\Delta E_G = \langle\Psi|\hat{H}'|\Psi\rangle = \langle 0|\hat{c}_{\mathbf{k}s}\left[\int d^3r\, m\hat{\psi}^{\dagger}(r)\beta\hat{\psi}(r)\Phi_G\right]\hat{c}_{\mathbf{k}s}^{\dagger}|0\rangle. \tag{B13}$$

Here we have used the gravitational potential term in the Hamiltonian of Eq. (4) and inserted the definition of $\hat{\rho}_G(\mathbf{r})$ from Eq. (B2). Using the form of $\hat{\psi}(r)$ from Eq. (7) gives the relevant terms in $\hat{\rho}_G(\mathbf{r})$ as

$$\hat{\rho}_G = m\left(\sum_{\mathbf{p'},s'}\sqrt{\frac{mc^2}{E_{\mathbf{p'}}}}\Big[\hat{c}(\mathbf{p'},s')v^\dagger(\mathbf{p'},s')e^{i\mathbf{p'}\cdot\mathbf{r}/\hbar}\Big]\right)\beta\left(\sum_{\mathbf{p},s}\sqrt{\frac{mc^2}{E_{\mathbf{p}}}}\Big[\hat{c}^\dagger(\mathbf{p},s)v(\mathbf{p},s)e^{-i\mathbf{p}\cdot\mathbf{r}/\hbar}\Big]\right)$$

$$= m\left(\sum_{\mathbf{p},s}\sum_{\mathbf{p'},s'}\sqrt{\frac{mc^2}{E_{\mathbf{p'}}}}\sqrt{\frac{mc^2}{E_{\mathbf{p}}}}v^\dagger(\mathbf{p'},s')\beta v(\mathbf{p},s)e^{i\mathbf{p'}\cdot\mathbf{r}/\hbar}e^{-i\mathbf{p}\cdot\mathbf{r}/\hbar}\right)\left(-\hat{c}^\dagger(\mathbf{p},s)\hat{c}(\mathbf{p'},s')+\delta_{\mathbf{pp'},ss'}\right)$$

(B14)

Here the order of the operators $\hat{c}(\mathbf{p'},s')$ and $\hat{c}^\dagger(\mathbf{p},s)$ were interchanged using their anti-commutator. The term involving the Kronecker delta function is a constant (non-operator) that is independent of the state of the system and can be ignored [68]. (This procedure is also necessary in order to obtain the correct charge of a positron.) Inserting Eq. (B14) into Eq. (B13) gives

$$\Delta E_G = -\left(\frac{mc^2}{E_{\mathbf{k}}}\right)\left(v^\dagger(\mathbf{k},s)\beta v(\mathbf{k},s)\right)m\Phi_G = -\left(\frac{mc^2}{E_{\mathbf{k}}}\right)\left(\tilde{v}(\mathbf{k},s)v(\mathbf{k},s)\right)m\Phi_G = \left(\frac{mc^2}{E_{\mathbf{k}}}\right)m\Phi_G. \quad \text{(B15)}$$

The right hand side of Eq. (B15) follows from the fact that $\tilde{v}(\mathbf{k},s)v(\mathbf{k},s)=-1$ [12,14,15].

It can be seen from Eq. (B15) that a positron will have a gravitational potential energy with the correct sign but multiplied by a relativistic factor that depends on the value of $\mathbf{k}$. This can be avoided if we define a new set of spinors $u'(\mathbf{k},s)$ and $v'(\mathbf{k},s)$ that are defined by

$$u'(\mathbf{k},s)=\sqrt{E_{\mathbf{k}}/mc^2}\,u(\mathbf{k},s)$$
$$v'(\mathbf{k},s)=\sqrt{E_{\mathbf{k}}/mc^2}\,v(\mathbf{k},s).$$

(B16)

We also define the operator $\hat{\psi}'(\mathbf{r})$ by

$$\hat{\psi}'(\mathbf{r})=\sum_{\mathbf{p},s}\sqrt{\frac{mc^2}{E_{\mathbf{p}}}}\Big[\hat{b}(\mathbf{p},s)u'(\mathbf{p},s)e^{i\mathbf{p}\cdot\mathbf{r}/\hbar}+\hat{c}^\dagger(\mathbf{p},s)v'(\mathbf{p},s)e^{-i\mathbf{p}\cdot\mathbf{r}/\hbar}\Big]$$

(B17)

where the spinors of Eq. (B16) have been inserted into Eq. (7). The definition of $\hat{\rho}_G(\mathbf{r})$ in Eq. (B2) is now replaced by

$$\hat{\rho}_G(\mathbf{r})=m\tilde{\psi}'(\mathbf{r})\hat{\psi}'(\mathbf{r})=m\hat{\psi}^\dagger{}'(\mathbf{r})\beta\hat{\psi}'(\mathbf{r}),$$

(B18)

while the usual field operator $\hat{\psi}(\mathbf{r})$ is used in $\hat{\rho}_E(\mathbf{r})$ and $\hat{\mathbf{j}}_E(\mathbf{r})$.

With this choice of $\hat{\rho}_G(\mathbf{r})$, Eq. (B15) becomes

$$\Delta E_G = m\Phi_G$$

(B19)

and the energy of a positron in a gravitational potential is $E_k + m\Phi_G$ as desired. The same result can also obtained for the energy of an electron. This justifies the assumption in the text that a virtual electron-positron pair has a gravitational potential energy of $2m\Phi_G$.

The nonrelativistic limit is not affected by this new definition of $\hat{\rho}_G(\mathbf{r})$ because $\hat{\psi}'(\mathbf{r}) \to \hat{\psi}(\mathbf{r})$ in that limit and Eqs. (B9) and (B12) remain valid. It can also be shown that the expectation value in the state $|\Psi\rangle$ of the integral of $\hat{\rho}_G(\mathbf{r})$ over all space is equal to the mass $m$ as would be expected.

It should be emphasized that the model described in this paper is only intended to provide an alternative and approximate description of the propagation of photons in a gravitational potential; it is not intended to represent a complete or consistent theory. For example, $\hat{\rho}_G(\mathbf{r})$ cannot obey a conservation law as a result of pair production and it is not part of a covariant four-vector. A more rigorous discussion of related issues in the currently-accepted formulation of the Dirac equation in curved spacetime will be submitted for publication elsewhere.